\documentclass[twocolumn,superscriptaddress,preprintnumbers,amsmath,amssymb,prx]{revtex4-1}

\newcommand{\IMSS}{Muon Science Laboratory and Condensed Matter Research Center, Institute of Materials Structure Science, High Energy Accelerator Research Organization (KEK-IMSS), Tsukuba, Ibaraki 305-0801, Japan}
\newcommand{\Tohoku}{Institute for Materials Research, Tohoku University (IMR), 2-1-1 Katahira, Aoba-ku, Sendai 980-8577, Japan}
\newcommand{\Ibaraki}{Graduate School of Science and Engineering, Ibaraki University 2-1-1 Bunkyo, Mito, Ibaraki, 310-8512, Japan}
\newcommand{\Sokendai}{Department of Materials Structure Science, The Graduate University for Advanced Studies (Sokendai), Tsukuba, Ibaraki 305-0801, Japan}
\newcommand{\NIMS}{National Institute for Materials Science (NIMS), Tsukuba, Ibaraki 305-0044, Japan}

\usepackage{txfonts}
\usepackage[dvipdfmx]{graphicx}
\usepackage{dcolumn} 
\usepackage{bm} 
\usepackage{xspace}
\usepackage{ulem} 
\usepackage{multirow}
\usepackage{tabularx}
\usepackage{breqn}
\usepackage[hypertex,colorlinks=true,linkcolor=black,citecolor=blue,filecolor=blue,urlcolor=blue,setpagesize=false,nesting=true]{hyperref}
\begin{document}

\author{H.~Okabe}
\thanks{Corresponding author: hirotaka.okabe.b4@tohoku.ac.jp}
\affiliation{\IMSS}\affiliation{\Tohoku}
\author{M.~Hiraishi}
\affiliation{\IMSS}\affiliation{\Ibaraki}
\author{A.~Koda}
\affiliation{\IMSS}\affiliation{\Sokendai}
\author{Y.~Matsushita}
\affiliation{\NIMS}
\author{T.~Ohsawa}
\affiliation{\NIMS}
\author{N.~Ohashi}
\affiliation{\NIMS}
\author{R.~Kadono}
\thanks{Corresponding author: ryosuke.kadono@kek.jp}
\affiliation{\IMSS}\affiliation{\Sokendai}


\title{Nano-scale dynamics of hydrogen in VO$_{2}$ studied by $\mu $SR}
\begin{abstract}
Hydrogen dynamics in the nanoscale region of VO$_{2}$ was investigated by muon spin rotation/relaxation ($\mu $SR) technique. Positively charged muon acts as a light radioisotope of proton and can be used as a probe to explore the inside of materials from an atomic perspective. By analyzing the muon hopping rate and the spatial distribution of the $^{51}$V nuclear magnetic moments, we have identified two types of diffusion paths in VO$_{2}$ (via oxygen-muon bonds or defects) and the potential of a high diffusion coefficient in the 10$^{-10}$ cm$^{2}$/s range at ambient temperature. Our results provide valuable information for developing hydrogen-driven VO$_{2}$--related electronic devices.
\end{abstract}

\maketitle

\section{Introduction}
Vanadium dioxide (VO$_{2}$) has drawn the interest of numerous researchers due to its pronounced thermal-driven metal-insulator (M-I) transition at  $T_{\rm MI}$ (= 340 K) with an enticing V-V dimerization in the low-temperature phase~\cite{Morin}. In parallel with the fierce debate on the M-I transition mechanism, much effort has been attempted to explore its device applications~\cite{Jero,Cav}. Since the resistance varies over several orders of magnitude across the M-I transition and the transition temperature can be tuned over a wide range by impurity doping, various applications are being explored. One of the applications is resistive random access memory (ReRAM)~\cite{Sawa}, which has the potential to realize next-generation nonvolatile memory devices with high-speed operation, high current density, and high-density recording by using VO$_{2}$ as a selector~\cite{Son}. Furthermore, ReRAM may apply to artificial neural networks due to its multi-level, analog memory, and hysteretic current-voltage (I-V) characteristics~\cite{Waser}.
\par
However, in recent years, Ji et al. pointed out that electrical switching behavior in VO$_{2}$-based electric double-layer transistors (EDLT) arises from the electrochemically driven doping of hydrogen ions~\cite{Ji}. It has been reported that impurity hydrogen in the ionic liquid used for the EDLT electrode migrates into VO$_{2}$ by the electric field and provides carriers, causing a change in resistance~\cite{Zhou,Shyb}. In general, hydrogen readily penetrates metal oxide semiconductors and is a possible source for unintentional carrier doping~\cite{Wale1,Wale2}. It has also been found that a considerable amount of hydrogen can reversibly enter and desorb from the VO$_{2}$ film, significantly affecting its electrical properties~\cite{Andre,Yoon}. Hydrogenation of VO$_{2}$ also induces structural modification; high hydrogen content leads to orthorhombic HVO$_{2}$ where intercalated hydrogen accommodates in an oxygen (O)-channel, forming typical OH bonds~\cite{Filin}. Therefore, the physical properties of VO$_{2}$ are strongly affected by hydrogen, and information on hydrogen dynamics is crucial for developing these devices.
\par
For hydrogen dynamics in VO$_{2}$, diffusion coefficients estimated from optical observations of the metal-insulator domain boundary and from measurements of transient electronic transport properties during proton intercalation have been reported~\cite{Lin,Mura}. Since these reports focus on the micron-order VO$_{2}$-HVO$_{2}$ transition process, a proton-relay hydrogen diffusion mechanism (Grotthuss mechanism~\cite{Gro}) with a high hydrogen concentration seems dominant. In contrast, VO$_{2}$ in electronic devices is typically a thin film of several tens of nm, and therefore the hydrogen content is small and may have a different diffusion mechanism than the above. However, observing the dynamics of trace amounts of hydrogen in devices is extremely difficult and remains unexplored.
\par
Here, we report on the dynamics of hydrogen at a dilute concentration in VO$_{2}$ investigated by muon spin rotation ($\mu $SR) spectroscopy. Positively charged muon ($\mu ^{+}$) is a subatomic particle that acts as a light radioisotope of proton, simulating the local state of hydrogen in the matter on an atomic scale while simultaneously probing one's own evironment~\cite{Seeg}. In addition, muons are injected directly into the material as a pulsed ion beam generated by an accelerator (about 10$^{3}$/cm$^{2}$ per pulse), so there is no need to fabricate the sample into a thin film or device. The clarification of hydrogen dynamics in VO$_{2}$ in the nanoscale region using muons is expected to provide valuable insights into the development of next-generation electronic devices such as ReRAM and hydrogen-driven devices.
\par
\section{Experimental Details}
We used a powder sample (M180 $\mu$m pass) of VO$_{2}$ (99.9\% purity, Kojundo Chemical Lab. Co., Ltd.). The crystal structure and phase purity were checked using the powder X-ray diffractometer (SmartLab, Rigaku Co.) from 300 to 400 K (see the Supplemental Material, Fig. S1-S3). The structural transition from the low-temperature orthorhombic phase to the high-temperature tetragonal (rutile) phase via the intermediate phase was observed around 340 K, as reported previously~\cite{Kim}. Considering the possibility that the sample already contained hydrogen, the hydrogen content was observed by thermal desorption spectrometry (TDS1200, ESCO) (Fig. S4). The amount of desorbed hydrogen was 1.14$\times$10$^{19}$ cm$^{-3}$ (corresponding to H$_{0.00034}$VO$_{2}$) upon increasing the temperature to 1073 K. The magnetic susceptibility was measured using the SQUID magnetometer (MPMS, Quantum Design, Inc.) from 5 to 400 K. No magnetic anomalies were observed in this temperature range, except for the M-I transition of VO$_{2}$ [Fig. 2(a)]. Vanadium oxides have many analogs with slightly different oxygen ratios that exhibit different magnetic transition temperatures~\cite{CVD}. Based on these results, our sample quality is adequate for studying the nature of VO$_{2}$. $\mu $SR experiments were performed using the ARTEMIS spectrometer installed in the S1 area at Muon Science Establishment (MUSE), Japan Proton Accelerator Research Complex (J-PARC).
\section{Results and discussion}
Figure 1(a) shows zero-field $\mu $SR time spectra at several temperatures from 5 K to 380 K. These spectra above $\sim$200 K show the Gaussian-like line shapes mainly attributed to the random local fields from the nuclear magnetic moment of $^{51}$V (5.148 $\mu_{\rm N}$). However, below 150 K, it can be seen that a fast exponential relaxation with partial loss of asymmetry appears in the initial time region ($t\leq\sim$1 $\mu $s). This suggests the gradual development of microscopic magnetic order of vanadium electron spins in parts of the sample, which may result from lattice imperfections ~\cite{LT29}.
\par
\begin{figure}[!t]
	\centering
	\includegraphics[width=\linewidth,clip]{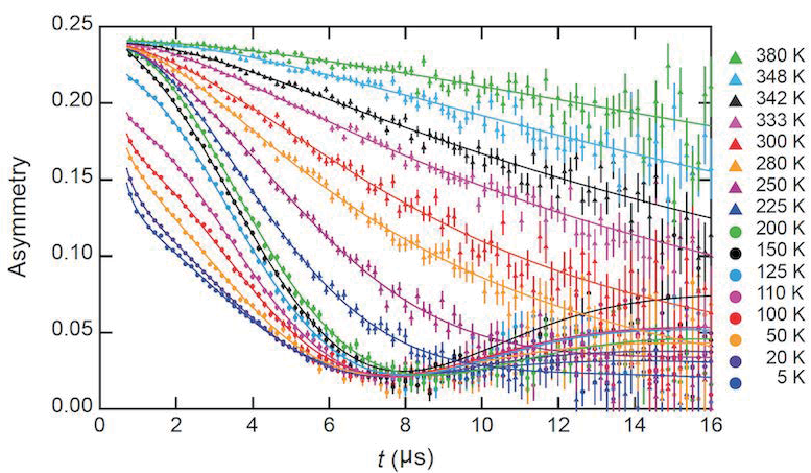}
	\caption{Zero-field $\mu $SR time spectra in VO$_2$ measured at various temperatures (5 K$\sim$380 K). The solid lines represent the result of least-squares fitting by Eq.1. All spectra were measured in the temperature increasing process.}
	\label{Fig1}
\end{figure}
Based on the above observation, we analyzed the time spectra using the following curve-fit function:
\[G_{\mathrm{Z}}(t)=AG_{\mathrm{KT}}(\Delta,\nu,B_{\mathrm{ext}},t)e^{-\lambda t}+A_{\mathrm{bg}}.\ \ (1)\]
\noindent
A is a signal amplitude called asymmetry, which reflects the volume fraction of the sample. The first term on the right-hand side is the product of the dynamic Gaussian Kubo-Toyabe function $G_{\rm{KT}}(\Delta,\nu,B_{\rm{ext}},t)$~\cite{Haya} which represents the relaxation due to random local fields exerted from nuclear magnetic moments (described by nuclear dipole-field width $\Delta$, hopping rate of the muon $\nu$, longitudinal field $B_{\rm{ext}}$), and the exponential relaxation component due to hyperfine interaction with electron spins (relaxation rate $\lambda$). The second term on the right-hand side ($A_{\rm{bg}}$) represents the $T$-independent background term from the surrounding sample holder and cryostat walls. When $\Delta \geq \nu$, the muon spin depolarization can be quenched by applying $B_{\rm ext}\approx$ 2--3$\times\Delta/\gamma_\mu$, so a global fit of the spectra at different $B_{\rm{ext}}$ (0, 5, 10 G) using Eq. (1) can determine $\Delta$ and $\nu$ reliably at each temperature (Fig. S5). The longitudinal field dependence of the time spectrum is clearly observed even after 10 $\mu$s. This fact suggests that the high-flux pulsed muon at J-PARC has enabled one to separate $\nu$ and $\Delta$ in VO$_{2}$ clearly for the first time. \color{black}
\par
\begin{figure}[!t]
	\centering
	\includegraphics[width=\linewidth,clip]{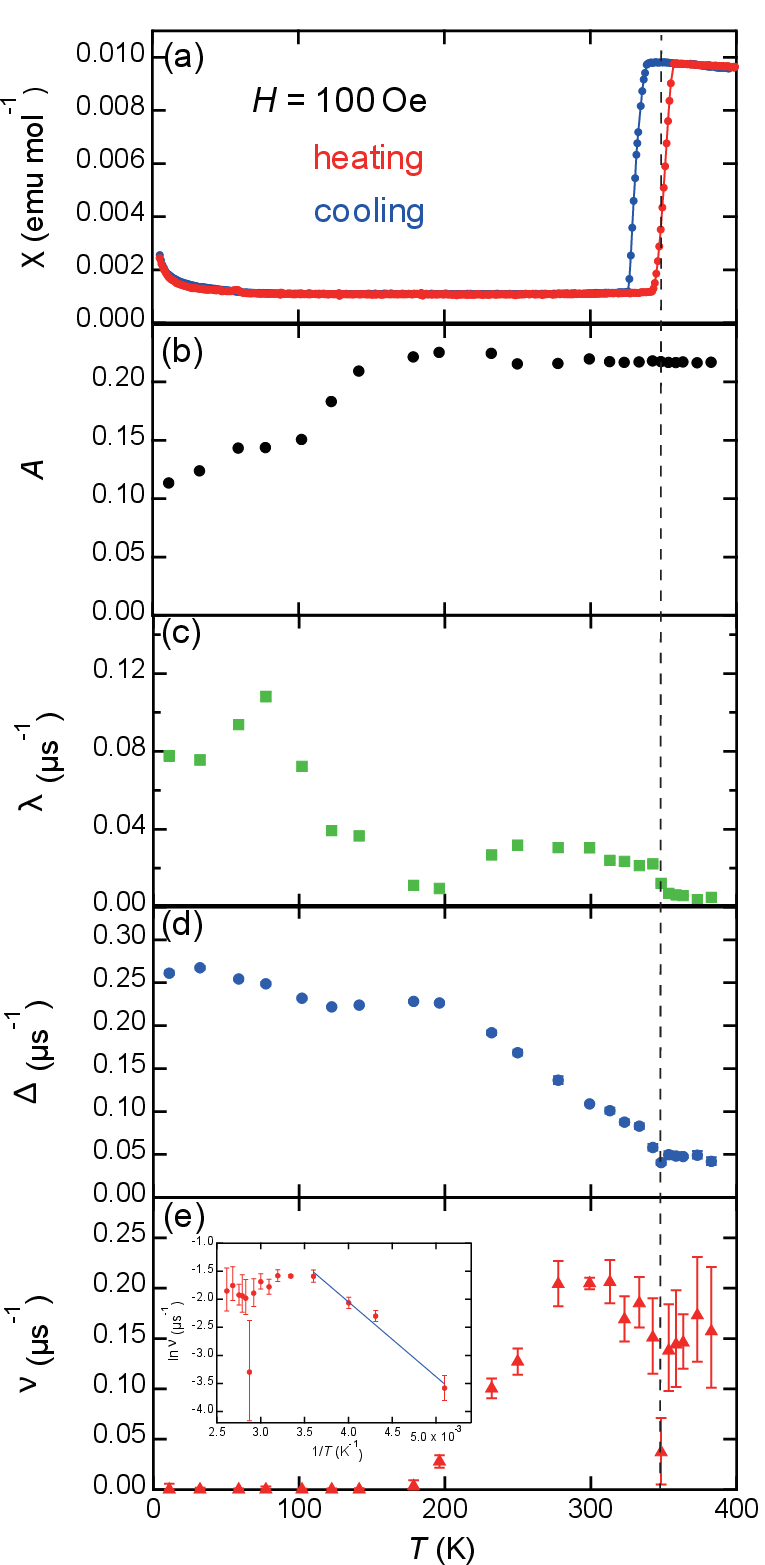}
	\caption{Temperature dependence of (a) the magnetic susceptibility $\chi$, (b) the asymmetry $A$, (c) the relaxation rate $\lambda$, (d) the nuclear dipole-field width $\Delta$, and (e) the hopping rate $\nu$. The inset in (e) shows the Arrhenius plot of $\nu$ as a function of 1/$T$ with a line of best fit from 200 K to 280 K. The temperature dependence of $\nu$ is well explained by a thermal activation process, $\nu = \nu_{0}$exp(-$E_{a}/kT$), where $\nu_{0}[$=24(3) $\mu$s$^{-1}$] is the prefactor (pre-exponential) dependent on the dipole-dipole coupling between muon and nuclear spins, and $E_a$[=0.11 (1) eV] is the activation barrier for muon diffusion.}
	\label{Fig2}
\end{figure}
Figures 2 shows the temperature dependencies of the magnetic susceptibility $\chi$ and the $\mu $SR fitting parameters $A$, $\lambda$, $\Delta$ and $\nu$. $\chi$ exhibits a paramagnetic-nonmagnetic transition at  $T_{\rm MI}$ with a thermal hysteresis of 10 K, while there is no indication of any magnetic anomalies below $T_{\rm MI}$ [Fig. 2(a)]. While this is seemingly inconsistent with the decrease of $A$ (reflecting the volume fraction of the nonmagnetic component) and associated increase in $\lambda$ below $\sim$150 K (indicating the emergence of the local field from unpaired electron spins) [Fig. 2(b) and (c)], our previous study indicates that this is due to the coexistence of the singlet state and microscopic magnetic ordered states caused by structural imperfections, such as lattice defects~\cite{LT29}. The increase in $\lambda$ above 180 K is due to two possible reasons: (1) unpaired electrons are located adjacent to the defect where muons are trapped, forming a polaron-like state~\cite{Shimop,Vilaop,Itop,Dehnp1,Dehnp2}, and (2) $\lambda$ and nuclear relaxation are not completely separated ($\lambda$ is affected by changes in $\nu$) in the high-temperature region where $\nu\gg\Delta$. (A discussion of magnetism is beyond the scope of this paper and will be omitted.) 
\par
$\Delta$ shows a gradual decrease with increasing temperature, markedly above 200 K [Fig. 2(d)]. Note that the spectra could not be reproduced by Eq. (1) with $\Delta$ fixed to a constant (Fig. S6). This is not expected unless the $^{51}$V distribution varies due to changes in crystal structure or muon sites [see Eq. (2) below]. Considering that the fluctuation rate {$\nu$} increases in the relevant temperature range [Fig. 2(e)], the decrease in $\Delta$ can be attributed to the change of the muon site by diffusion; it is unlikely that the diffusion of V ions ($\sim$500 times heavier than muons) sets in at such low temperatures. The activation energy $E_{\rm a}$ of the hopping motion estimated from the slope of the line in the Arrhenius plot for $\nu$ the inset of Figs. 2(e), is approximately 0.11 eV in the temperature range from 200 K to 280 K. This value is comparable to the activation energy for hydrogen diffusion (0.08-0.12 eV) in Sr/BaTiO{$_{(3-x)}$H$_{x}$ hydride conductors~\cite{Ito,BTO}. The increase in $\nu$ begins to be observed around 180 K and reaches about 0.15 $\mu$s$^{-1}$ around 250 K, which is similar to that reported for other oxides as a result of $\mu ^{+}$ diffusion ~\cite{Nishida,Sonier}. On the other hand, $\nu$ is suppressed in the high-temperature side above $\sim$250 K and saturated at 280 K. This also correlates with a decrease in $\Delta$, suggesting that the diffusion pathway may change near the structural phase transition. Thus, the pre-exponential coefficients obtained by fitting the data, including 280 K with the Arrhenius formula, are about an order of magnitude smaller than for other oxides, which also supports such a change in the diffusion mechanism. \color{black}
\par
\begin{figure}[!t]
	\centering
	\includegraphics[width=\linewidth,clip]{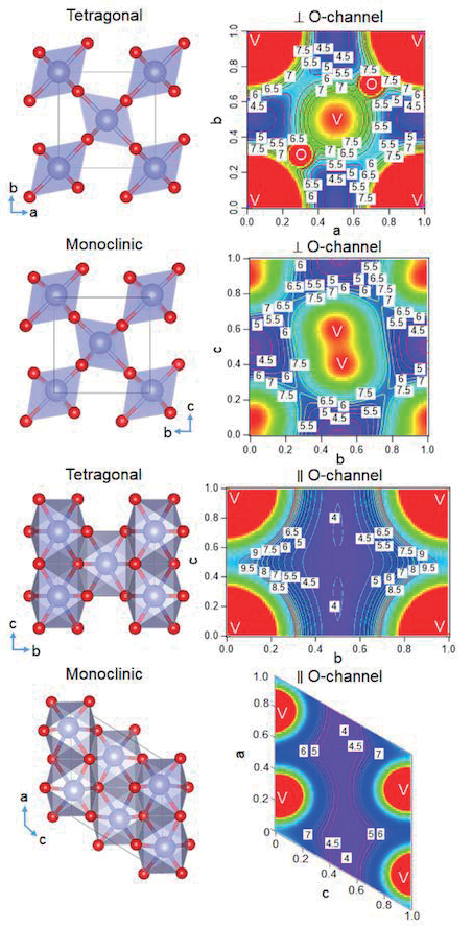}
	\caption{Calculated distribution of $\Delta$ for the cross-sectional layers parallel or perpendicular to the oxygen channels in the tetragonal and the monoclinic VO$_2$. The values of $\Delta$ (numerical values in the contour map, unit: Gauss) are calculated by using Dipelec code~\cite{Koji} with reported lattice parameters~\cite{Eyert}. Crystal structures were visualized using the VESTA program~\cite{Izumi}.}
	\label{Fig3}
\end{figure}
Above room temperature, the increase in $\nu$ shows a sharp drop around $T_{\rm MI}$ [Fig. 2(e) and Fig. S7(e)]. This behavior may attribute to the trapping effect by lattice defects and phase boundaries. The muons, which diffuse rapidly due to the temperature rise, are easily trapped in the lattice defects that require considerable activation energy to escape~\cite{Vilao}. Furthermore, since three phases coexist during the M-I transition of VO$_{2}$, the distribution of  $E_{\rm a}$ is expected to be larger at the phase boundary due to structural disorder associated with tweedy or glassy textures of mixed phases with characteristic correlation lengths of $\sim$5 nm observed in VO$_{2}$ epitaxial films~\cite{twd}. In addition, the carrier mobility measurement has inferred a strong scattering of electrons at the boundaries of metallic and insulating domains in the inhomogeneous transition of VO$_{2}$~\cite{Fu}. Given the above facts, the strong lattice inhomogeneity in the coexisting state inherent to the first-order phase transition contributes to the sharp drop of $\nu$ around $T_{\rm MI}$. This is also supported by the fact that the dropping temperature of $\nu$ roughly coincides with the appearance of the coexisting state (Fig. S7). 
\par
To discuss the crystallographic sites of muon (or hydrogen) in VO$_{2}$, we performed the theoretical calculation of $\Delta$ versus muon position. Since muon is an ultra-high sensitivity probe of local magnetic fields, we can estimate the muon sites using a spatial configuration of $^{51}$V nuclear magnetic moments as a guide. In polycrystals under zero magnetic field, $\Delta^{2}$ can be evaluated as the second moment of internal field distribution exerted from nuclear magnetic moments ($I$ = 7/2 in $^{51}$V, which is subject to quadrupole interaction) using the following formula~\cite{Hemp}: 
\[\Delta^{2} \approx \frac{4}{9}\gamma_{\mu}^{2}\left(\frac{\mu_{0}}{4\pi}\right)^{2}\sum\frac{\mu_{i}^{2}}{r_{i}^{6}},\ \ (2)\]
\noindent
where $\gamma_{\mu}$ (= 2$\pi \times $13.553 kHz/G) is the muon gyromagnetic ratio, $r_i$ is the distance from the $i$-th nuclear magnetic moment $\mu_{i}$, and $\mu_{0}$ is the vacuum permeability. Figure 3 shows the calculated spatial distribution of $\Delta$ for the cross-sectional layers parallel or perpendicular to the oxygen channels in the tetragonal and the monoclinic structures. Comparison of the calculated results for both structures shows that there is no significant difference in the $\Delta$ distribution within each oxygen channel.  The value of $\Delta$/$\gamma_{\mu}$ is just under 4 G ($\Delta$$\sim$0.34 $\mu$s$^{-1}$) at the center of the oxygen channel and has similar values along the channel direction. The maximum value of $\Delta$/$\gamma_{\mu}$ determined from experiments is about 3.2 G ($\Delta$$\sim$0.27 $\mu$s$^{-1}$) at 4 K, indicating that the muons are located near the center of the oxygen channel at low temperatures. This result is in interesting agreement with the muon/hydrogen case in $\beta$-MnO$_{2}$ which has a similar structure to VO$_{2}$ and is presumed to be a common property of rutile-type oxides~\cite{OKB}. 
\par
As the temperature increases, $\Delta$ decreases rapidly to about one-fifth of its low-temperature counterpart ($\Delta$$\sim$0.05 $\mu$s$^{-1}$) above $T_{\rm MI}$. The muon sites corresponding to such a small value of $\Delta$ are not found for the interstitial position, even considering a lattice distortion caused by hydrogen intercalation in the rutile structure. Therefore, it is highly likely that implanted muons are trapped in vacancies at high temperatures. Since the crystallite size of this sample is 50$\sim$80 nm (Fig. S3), the surface contribution is expected to be small. However, muons might have diffused to the grain boundaries at high temperatures considering the saturation of $\nu$ above 280 K. \color{black}
\par
We performed a simulation of the relationship between $\Delta$ and the size of the vacancy based on Eq. (2). The virtual element muon (Mu) was fixed at the center of the oxygen channel, and $\Delta$ was calculated for a variety of situations where the V ions at the first, second, and third nearest neighbors (V1,V2, and V3) were removed sequentially. Note that structural distortions due to the vacancy was not taken into account. Figure 4(a) shows the reduction rate of $\Delta$ as a function of the number of removed vanadium $N$. One can see that $\Delta$ decreases almost linearly as $N$ increases within the each V shell (V1-V3), down to about 20\% when the V ions in the V3 shell (the third nearest neighbors) are removed. Figure 4(b) compares the simulation result and the experimental $\Delta$. The existence of the vacancy of a radius of $\sim$0.38 nm (the distance between center Mu and V3) can explain the magnitude of $\Delta$ above $T_{\rm MI}$. The discrepancy in $\Delta$ between our result and that reported in the previous $\mu$SR study~\cite{Mengyan} can be interpreted as due to the difference in the oxygen ratios between samples. According to their $\mu$SR result, some magnetic order was observed even in the nonmagnetic state below $T_{\rm MI}$. While the details of the sample quality in Ref.~\cite{Mengyan} are unknown, magnetic secondary phases may have formed during the bulk sintering process. In any case, our results strongly suggest that muons at high temperatures spend most of their life in a diffusive process of repeated "capture and release" by vacancies. The $\mu$SR study in a similar trapping mechanism has been discussed in semiconductors~\cite{Hemp,Vilao}. These two different muons (interstitial and vacancy) also exist in $\beta$-MnO$_{2}$ and correspond to two other hydrogens called Ruetschi and Coleman protons in battery materials~\cite{Ruet1,Ruet2,Colm}.
\par
\begin{figure}[!t]
	\centering
	\includegraphics[width=\linewidth,clip]{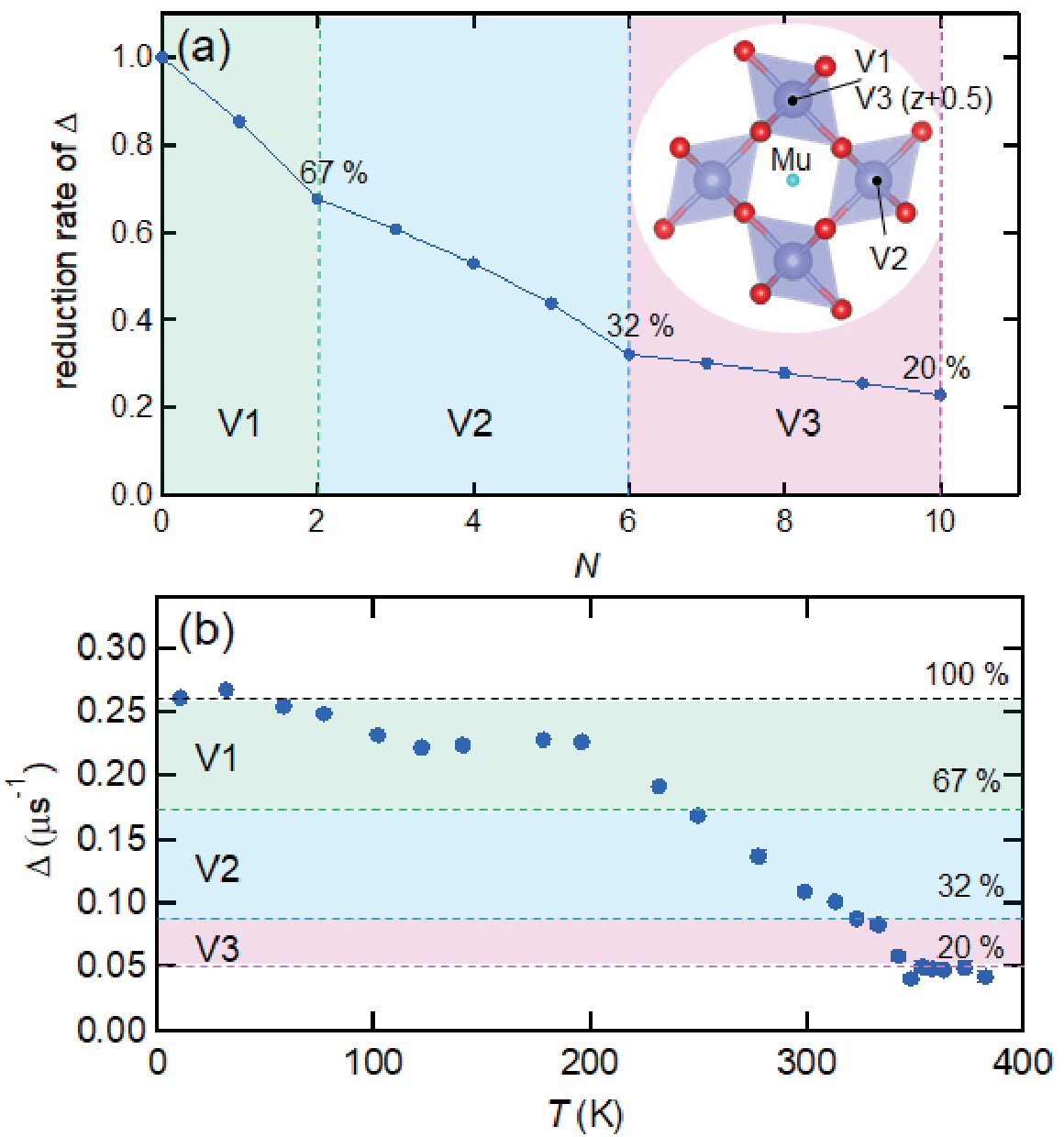}
	\caption{(a) The reduction rate of $\Delta$ as a function of the number of removed vanadium $N$. The inset shows the spatial arrangement of muon (Mu) and surrounding vanadium, the first (V1), second (V2), and third (V3) nearest neighbors. (b) Temperature dependence of the experimental $\Delta$ with the simulation result. The dotted lines indicate the reduction rate of $\Delta$ at each stage of removing V1, V2, and V3.}
	\label{Fig4}
\end{figure}
Finally, to gain further insight into the interstitial hydrogen dynamics in VO$_{2}$, we discuss muon diffusion in the oxygen channel at intermediate temperatures. The diffusion coefficient of muon $D_{\mu}$ is approximated by the following equation~\cite{Borg}:
\[D_{\mu} = \sum_{i=1}^{n}\frac{1}{N_{i}}Z_{i}s_{i}^{2}\nu,\ \ (3)\]
\noindent
where $N_{i}$ is the number of muon hopping sites in the $i$-th path, $Z_i$ is the vacancy fraction, and $s_i$ is the hopping distance. The interstitial hydrogen diffusion pathway is one-dimensional-like along the O-channel in VO$_{2}$, and the diffusion transverse to the channel can be negligible~\cite{John,Lin}. Assuming that muon diffusion via the O-Mu bond is dominant below room temperature, a muon jumps to one of the four neighboring oxygen atoms ($N_{i}$=4, the average O-O distance $s_i$=0.271 nm). Additionally, neglecting the existence of other muons or hydrogen ($Z_i$=1) and taking the experimental value of the hopping rate at 280 K [$\nu$=0.20(2) $\mu$s$^{-1}$] to Eq. (3) yields $D_{\mu}$=1.47(2)$\times$10$^{-10}$ cm$^{2}$/s. This value is on the same order of magnitude as the reported diffusion constants of hydrogen $D_{\mu}$=0.46$\times$10$^{-10}$ cm$^{2}$/s at 300 K~\cite{Mura} and $D_{\mu}$=6.7$\times$10$^{-10}$ cm$^{2}$/s at 373 K~\cite{Lin}, indicating that the result of our hydrogen simulation by muon is reliable; the mass difference (the muon's mass is 1/9th that of the proton) is unimportant at high temperatures where only moderate isotope dependences, e.g., $D_{\mu}$=4.7(9)$\times$10$^{-4}$ cm$^{2}$/s and $D_{\mu}$=7.5$\times$10$^{-4}$ cm$^{2}$/s above 300 K in iron~\cite{Gpan}, are expected. Interestingly, 
it is about three orders of magnitude larger than $D_{\mu}$ (=1.8$\times$10$^{-13}$ cm$^{2}$/s at 298 K) in TiO$_{2}$, which has the same crystal structure~\cite{John}. Based on the above results, we conclude that the interstitial hydrogen in VO$_{2}$ has the potential to diffuse at a sufficient rate (i.e., $\sim$10$^{-10}$ cm$^{2}$/s) even at room temperature, suggesting that it is suitable for applications requiring a fast response, such as ReRAM. However, high-quality VO$_{2}$ thin films are required to fabricate high-reliability devices because defect/vacancy-mediated muon diffusion is observed even below room temperature [Figure 4(b)].
\par
\section{Summary}
In summary, we have provided nano-scale information on the hydrogen dynamics in VO$_{2}$ using muon as pseudo-hydrogen. The temperature dependence of the muon hopping rate and the spatial distribution of the $^{51}$V nuclear magnetic moments indicates the existence of two types of muon diffusion (via O-Mu bonds and defects/vacancies) and that a high-quality thin film can have a diffusion coefficient of the order of 10$^{-10}$ cm$^{2}$/s around room temperature. Our findings will provide a useful criterion for designing hydrogen-driven electronic devices in the future.
\par
\section*{Acknowledgment}
This work was supported by the MEXT Elements Strategy Initiative to Form Core Research Centers, from the Ministry of Education, Culture, Sports, Science, and Technology of Japan (MEXT) under Grant No. JPMXP0112101001 and partially supported by the MEXT Program: Data Creation and Utilization Type Material Research and Development Project under Grant No. JPMXP1122683430 and JSPS KAKENHI (Grant No. 20K05312, 19H05819). The $\mu $SR experiments were performed under user programs (Proposal No. 2019MS02) at the Materials and Life Science Experimental Facility of the J-PARC. We also acknowledge the Neutron Science and Technology Center, CROSS for the use of MPMS in their user laboratories.


\end{document}